\documentclass[]{interact}

\usepackage{epstopdf}
\usepackage[caption=false]{subfig}

\usepackage[numbers,sort&compress]{natbib}
\usepackage{multirow}
\usepackage{amsmath} 
\usepackage{algorithm} 
\usepackage{algorithmic} 
\usepackage{tabularx} 
\usepackage{ragged2e} 
\usepackage{caption} 
\usepackage{xcolor}
\bibpunct[, ]{[}{]}{,}{n}{,}{,}
\renewcommand\bibfont{\fontsize{10}{12}\selectfont}
\makeatletter
\def\NAT@def@citea{\def\@citea{\NAT@separator}}
\makeatother

\usepackage{url} 

\theoremstyle{plain}

\theoremstyle{definition}

\theoremstyle{remark}

\begin{document}

\articletype{RESEARCH ARTICLE}

\title{Priority Based Inter-twin Communication in Vehicular Digital Twin Networks}

\author{
\name{ Qasim Zia\textsuperscript{a*}\thanks{CONTACT Corresponding author. Email: qzia1@student.gsu.edu} ,Chenyu Wang\textsuperscript{a**}, Saide Zhu\textsuperscript{b} and Yingshu Li\textsuperscript{a}}
\affil{\textsuperscript{a}Department of Computer Science, Georgia State University, Atlanta, GA, 30303; \textsuperscript{b}Cybersecurity analytics and operations, Penn State Berks,Reading, PA, 19610}
}

\maketitle

\begin{abstract}
With the advancement and boom of autonomous vehicles, vehicular digital twins (VDTs) have become an emerging research area. VDT can solve the issues related to autonomous vehicles and provide improved and enhanced services to users. Recent studies have demonstrated the potential of using priorities in acquiring improved response time. However, since VDT is comprised of intra-twin and inter-twin communication, it leads to a reduced response time as traffic congestion grows, which causes issues in the form of accidents. It would be encouraging if priorities could be used in inter-twin communication of VDT for data sharing and processing tasks. Moreover, it would also be effective for managing the communication overhead on the digital twin layer of the cloud. This paper proposes a novel priority-based inter-twin communication in VDT to address this issue. We formulate the problem for priorities of digital twins and applications according to their categories. In addition, we describe the priority-based inter-twin communication in VDT in detail, and algorithms for priority communication for intra-twin and inter-twin are designed, respectively. Finally, experiments on different priority tasks are conducted and compared with two existing algorithms, demonstrating our proposed algorithm's effectiveness and efficiency.
\end{abstract}

\begin{keywords}
 Vehicular digital twins, autonomous vehicles, vehicular Ad hoc networks, priority, reliability, vehicular communication, emergency messages.
\end{keywords}

\section{Introduction}

As communication technologies and the Internet of Things are making rapid progress, autonomous vehicles with computing, sensing, and communication have acquired attention from all over the world. People from academia and industry are investing more effort and time to improve its performance. Recent advancements in AI-driven regression models and feature extraction highlight the potential of data-driven approaches to enhance autonomous systems~\cite{ahmed2023robust}.  According to an estimate, it is projected that autonomous vehicles will account for 75\% of all new vehicles worldwide by the year 2040  \cite{su2018distributed}. There is no question about the uprising of smart cities and human transportation due to autonomous vehicles.

In this circumstance, the proposition of vehicular ad hoc networks (VANET) enables productive safety apps that inform the drivers of dangerous situations, for instance, accidents through broadcasting using a wireless medium. For traffic and safety efficiency, message dissemination is critical. Particular routing protocols \cite{sunit2019tutorial} are proposed for message dissemination in VANET and are considered according to the initiator's location, cluster-based, geocast, topology, platoons, etc. To disseminate the message rapidly, delays must be as minimal as possible. For this purpose, there is a strong desire for reliable and rapid message dissemination for emergencies and to decrease network load. This will help us in reducing accidents.

Digital Twins technology and VANET have led to a new era of intelligent transportation systems\cite{ali2023review}. Digital twins play a vital role in the resolution of autonomous vehicle issues. The vehicular digital twin (VDT) is interpreted as the digital representation of each instance of the physical vehicle on the cloud. It is constantly synchronized through wireless communication with the vehicle's real-time sensing data.  As shown in Fig.1, each autonomous vehicle uploads the vehicle status information and its private data to the digital twins over intra-twin communication\cite{sundar2014implementing}. However, autonomous vehicles have limited capacity to process data, and digital twins can process their enormous data using cloud computing {\cite{ali2023review}. The integration of advanced computational models further demonstrates the potential of sophisticated data processing in real-time systems~\cite{ahmed2025wallepp}.  {This is one of the reasons that Digital Twins are used in VANET more and more to improve predictive maintenance, autonomous driving, and real-time monitoring \cite{ali2023review}. In the digital twin layer, the digital twin among themselves can share their data through inter-twin communication \cite{sundar2014implementing} to help autonomous vehicles acquire real-time information and increase their range to other vehicle data. Due to the communication between digital twins, both within and between them, we can exchange information and data that would have been challenging to share at the physical layer which raises safety and efficiency standards \cite{guo2024survey}. 

However, handling massive quantities of data shared between digital twins is one of the main problems in the complexity of VDT\cite{he2022security}. To allow rapid and correct decision-making, particularly in situations of events related to emergency vehicles, priority communication between digital twins is vital\cite{ricci2022web}. Since not all vehicle services are as urgent or important as emergency vehicles are, so priority-based communication practices are clearly necessary in these situations\cite{martinez2010emergency}.  
Specifically, emergency vehicles will have to react rapidly to provide their emergency services. As a result, they should have free space on the roads and main highways. A slight delay in the transfer time might result in financial loss or the loss of someone's valuable life \cite{ghasemi2013stable}. It is essential to ensure the network channel's lowest time interval for emergency vehicles to reach their destination in time. The motivation for this effort is linked to the need for this essential importance. This includes creating ideal and rapid emergency services and mobility infrastructure in vehicular networks. We aim to decrease the message transmission time of a priority essential vehicle, such as a firefighting vehicle, from the point of source to destination.

\begin{table}[!t]
\caption{\textbf{Nomenclature}\label{tab:table1}}
\centering
\begin{tabular}{|c||c|}
\hline
\textbf{Symbol} & \textbf{Description}\\
\hline
$\mathcal{M}$ & The collection of vehicle’s DT\\
\hline
$m$ & A VDT  $m$ \\
\hline
$D_{DT}^{m} $
& The dataset in Vehicle’s DT $m$\\
\hline
$DT_{cat} $
 & The category of the Vehicle’s DT\\
\hline
$ App_{cat} $ & The category of the application\\
\hline
$P_{DT}$
& The priority of the Vehicle’s DT\\
\hline
$P_{app}$ 
 & The priority of the application\\
\hline
$PD$ & The private data of vehicle\\
\hline
$k$ & The $k$-th update in a round \\
\hline
$PI$ & {The private information sent to the vehicle} \\
\hline
\multirow{3}{*}{\shortstack{$V_{high}$\\$V_{medium}$\\$V_{normal}$}}  & \multirow{3}{*}{\shortstack{The high-priority vehicle, \\ The medium-priority vehicle,\\ and the normal-priority vehicle}}\\
&\\
&\\
\hline
\multirow{3}{*}{\shortstack{$DT_{high}$\\$DT_{medium}$ \\ 
$DT_{normal}$}} & \multirow{3}{*}{\shortstack{The high-priority vehicle’s DT,\\ The medium-priority vehicle’s DT,\\ and the normal-priority vehicle’s DT}} \\
& \\
& \\
\hline
\end{tabular}
\end{table}

\begin{figure*}[!ht]
    \centering
    \includegraphics[width=5.6 in]{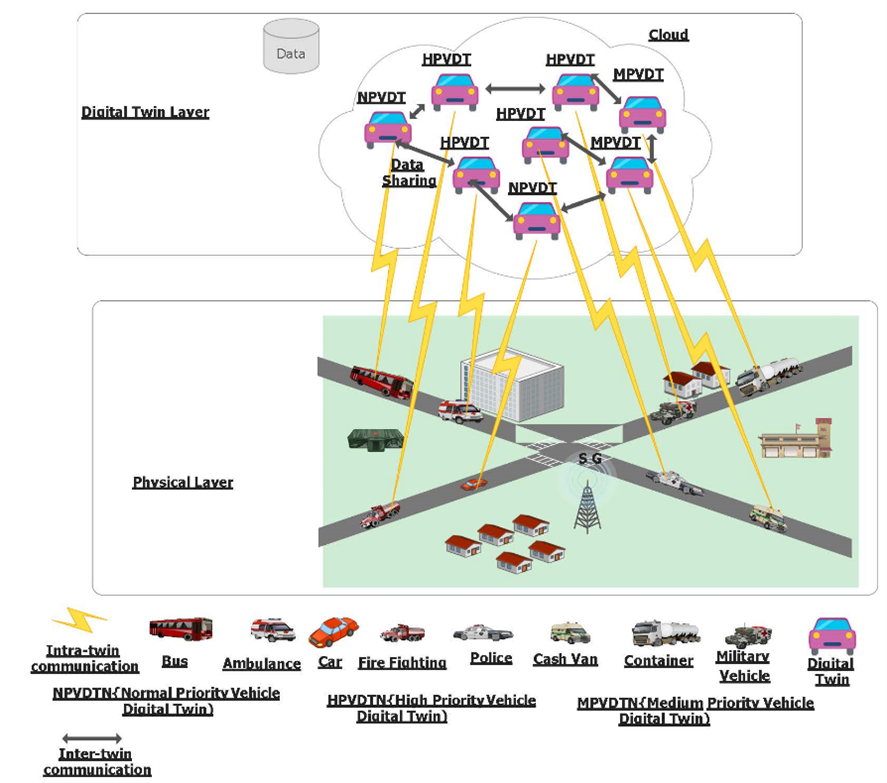}
    \caption{General architecture of priority-based vehicular digital twin networks}
    \label{fig_1}
\end{figure*}

We would essentially like to deliver messages to a vehicle that is an \textit{absolute priority}. To deliver the messages at the lowest possible time to an emergency vehicle, such as an ambulance, all accessible network potentials are utilized as an absolute priority. The data processing and sharing are prioritized depending on their significance for timely transmission. However, there are some of the most common issues, even at vehicular digital twin networks (VDTN), when it comes to processing and sharing data:
\begin{itemize}
\item \textbf{Same processing preference}: the processing preference of VDTs of autonomous vehicles is the same for every digital twin instance at the digital twin layer. 
\item \textbf{Same sharing preference}: the data-sharing preference among VDTs of autonomous vehicles is the same for every digital twin instance at the digital twin layer.
\item \textbf{Substantial communication overhead}: there would be a communication bottleneck as the number of vehicles that cooperate further increases because cloud computing cannot timely meet the concurrent requests from an incredible number of vehicles.
\end{itemize}
As a result, priority-based processing \cite{zia2016improving} and sharing of messages and data will be an essential part of this general architecture at the digital twin layer as vehicles become increasingly intelligent. Recently, VANET has been proposed to provide many road safety services \cite{hussain2017efficient}. Taking VANET as an example, in the real world, if firefighting vehicle services require some processing and information gathering from other vehicles, data should be processed and gathered as soon as possible to eliminate any potential delays. In this case, priority should be given to the digital twins of these vehicles because any minor delay in data processing could result in the loss of life especially during high-traffic hours e.g., school, office start and end time, or at the end of sports matches. Any delay presents a significant issue in the context of VDTs. Consequently, the \textit{priority} of different digital twin instances must be taken into consideration while allocating cloud computing resources \cite{ghanbari2012priority}.

Our work is the first one to review and implement priorities for the digital twins at the digital twin layer of the VDT architecture in contrast to the earlier work where priorities are implemented at the physical layer. An emergency vehicle through digital twins can broadcast its emergency message to a wide range compared to vehicles at the physical layer, with less range to broadcast its messages due to limited sensing capability. Unlike the previous work where all vehicles' digital twins are treated with the same precedence, our work assigns priorities to the digital twins of the vehicles according to the categories of vehicles. Through the use of this algorithm, we expect to increase the consistency of vehicular digital twins for emergency services, enhance network performance, and eventually provide an effective and safe emergency transportation system\cite{feng2023resilience}. Our algorithm is scalable in the dynamic environment and allocates the resources according to the highest priority vehicles available at a given time. Additionally, we offer a simple VDT-based architecture. We fully utilize this advantage to improve response time, and the main contributions and components related to our work are summarized as:
\begin{enumerate}
\item First, we formulate a problem definition for our VDTN, which provides a basis for designing the architecture;
\item Second, we design a priority-based VDTN to reduce communication overhead and improve response time.
\item Third, a comprehensive series of simulation experiments are carried out to assess how well the suggested priority-based inter-twin communication in the VDT framework, executes for the performance metrics. Our results demonstrate higher effectiveness, scalability, and efficiency. 

\end{enumerate}

The structure of this paper is as follows.
We first introduce the related work in Section \ref{sec:related_work}. Next, we present the problem formulation in Section \ref{sec:prob_form} and discuss the architecture of the priority-based VDTN in Section \ref{sec:priority_based_comm}. Related evaluation and results are presented in Section \ref{sec:eval}. Finally, we conclude the paper in Section \ref{sec:concl}  and plan possible guidelines for future research.

\section{Related work} \label{sec:related_work}
The VDT development has achieved growing interest in improving traffic efficiency and road safety\textcolor{brown}{\cite{almeaibed2021digital}}. For this purpose, communication between digital twins plays an important role in obtaining these objectives. Various studies have been performed to examine VDTN’s communication architectures. However, all of them have delivered data transmission without priority information exchange consideration. In this section, we briefly review the state-of-the-art studies focusing on VDT as well as their improvement.

\subsection{Priority in VANET}
Hussain \textit{et al.} \cite{hussain2017efficient}  proposed an RSU-based efficient channel access scheme that considers the subsequent parameters in order, namely, the speed of the vehicle, the direction of the vehicle to RSU, and services related to the emergency. The initial two parameters are utilized for each vehicle to determine the value of the earliest deadline first (EDF) while the third parameter is needed to break ties so that the first served vehicle would be an emergency vehicle if a similar value of EDF is there for many vehicles. In this scenario, low or moderate EDF-value vehicles are considered to be of high priority, while low-priority vehicles are those with high EDF values.

Tian \textit{et al.} \cite{tian2019priority} proposed a novel plan of priority assignment that aims to decrease transmission time. Based on the dynamic and static properties as well as the message size, each transmission was prioritized. The static factor is defined by the application type and message content. Dynamic factors are decided by the conditions of VANET in difference to static variables, specifically consisting of message usefulness, the recipient and sender directions of vehicles, geographic position, vehicle velocity, message validity, and meteorological conditions. This work decides on message priority by utilizing unique metrics for geographical location and instance weather, which finally determine a dynamic scheduling order.

\subsection{Emergency vehicles in VANET}
In particular situations, every second matters, especially for emergency services like ambulances while in motion. Shirani \textit{et al.} \cite{shirani2009absolute} have presented an infrastructure-free approach that makes sure that an emergency vehicle gets uninterrupted passage by establishing control over traffic lights and the shortest possible route. VANET is one solution to decrease travel time. To find the most efficient route, the movement and speed of all other vehicles should be considered along with the destination of the vehicle. 

The technique discussed in \cite{padiyafair} utilizes a method that measures the distance between the intersection and the emergency vehicle during an urgent situation. It then informs drivers about the coming emergency vehicle arrival, which allows vehicles with higher priority to pass through the intersection before low-priority vehicles. The main objective of this system is to enhance reliability and safety while decreasing the time for life-saving vehicles to reach junctions. In the proposed method, the Euclidean distance formula is used to decide the shortest distance between emergency vehicles and junctions, and the vehicles with the shortest distance are given more priority to go through the intersection.

\subsection{Protocols and algorithms in VANET}

Ameddah \textit{et al.} \cite{ameddah2018priority} used a distinctive priority algorithm to regulate traffic through vehicle-to-vehicle communication at junctions. Vehicles establish coalitions based on priority, and the algorithm allows higher-priority vehicles to pass through the intersection before low-priority vehicles. However, vehicles may come across many crossroads while moving to the destination. There might be various routes to the same location. A vehicle’s wait time at the junction is calculated based on the priority as well as the load of all the vehicles to that junction’s entrance. Many factors such as priority, distance from the junction, vehicle direction, and density are considered during the waiting time calculation. In \cite{tantaoui2020real}, the authors have presented using Big Data a real-time prediction system to enhance the VANET network. They initially calculate average speed and traffic density at every section of the road and then through parallel processing, they predict the vehicle accident risk because of which there is faster execution.

In \cite{tian2018distributed}, Tian \textit{et al.} proposed a location-based protocol for broadcasting emergency messages in VANETs to reduce traffic accidents. When a vehicle is engaged in an accident, it directs a warning signal to other vehicles, assisting them in making informed decisions to prevent the accident zone. The proposed method helps to eliminate unnecessary rebroadcasting of emergency signals and can be utilized not only on roads but also in urban settings due to its broadcasting nature. The protocol effectively delivers communications to specific locations with minimal delivery latency and lag. It also eliminates undesired retransmissions and avoids emergency message broadcast collisions. Nevertheless, keeping real-time data can be difficult because of message collisions and the rapid exchange of beacon signals. In \cite{choi2019joint}, the authors prioritized messages to improve congestion for crucial messages.

Suthaputchakun \textit{et al.} \cite{suthaputchakun2013trinary} suggested a multi-hop message broadcast approach in VANET for time-critical emergency services. The method suggests a ternary partitioned black-burst-based broadcast protocol (3P3B) that comprises a micro-distributed interframe space (DIFS) framework at the MAC layer to prioritize emergency signals in communication channels for time-critical distribution over other communications, and a ternary partitioning approach divides the communication range into smaller parts. The main aim of this method is to enable the sender node in the distant sector of the communication range to send emergency messages, thereby increasing the speed of message dissemination by decreasing the number of hops to the destination. Furthermore, the proposed protocol extensively reduces jitter in the contention window causing a stable contention period despite traffic volume. However, this approach has the disadvantage of substantial network latency because of heavy competition and concurrent transmissions. The authors in \cite{abdou2015priority} discuss a novel Autonomic Dissemination Method (ADM) for sending messages based on density levels and priority. Meanwhile,  Du \textit{et al.} in \cite{du2020vehicle} have discussed an adaptive backoff algorithm that depends on the number of vehicle nodes, but it does not consider the message priority. Based on message size, message priority is determined in \cite{tian2019improved}.  

\subsection{Architecture of VDTN}
In \cite{he2022security}, the author begins by providing a detailed overview of the architecture of the VDTN and looks at some of the applications related to VDT. It then discusses privacy and security challenges related to VDT. In the end, the paper concludes by suggesting several potential solutions and emphasizing some of the open research issues related to the privacy and security of VDT.  Based on the analysis of the literature related to vehicles and message prioritization, it is understood that limited research has been offered on message prioritization in digital twins. This could be because of the difficulty involved in defining the priorities of the vehicles and their associated digital twins, which could increase processing time. 

\section{Problem Formulation} \label{sec:prob_form}
In this section, we present the formulation of priority-based inter-twin communication in VDTN with vehicle data within its two-layer structure, providing a foundation for the algorithm and protocol design. We will discuss datasets, data, and priority representation concepts which will later be used in the algorithm.

Assume that in a two-layer VDTN as discussed above, there is a set of VDTs, i.e., $\mathcal{M}$, in the digital-twin layer, where each VDT shares data in the cloud. To calculate the number of VDTs in a set, a function $|\cdot|$ is used, e.g., $|\mathcal{M}|$. The total dataset generated by VDT in the Digital Twin Layer is represented by
$$ \sum_{i=1}^nD_{DT}^{i} $$
Suppose there is x data transit rate. To calculate the total amount of data that can be served in the given time T is xT. Thus, the following equation gives the total data that can be served in a given time T. 
$$ \sum_{i=1}^nD_{DT}^{i}\leq xT $$

The local dataset of a VDT $m$ can be stated as $$D_{DT}^{m}:=\{S_{n}^{m}\}_{n=0}^{\left|D_{DT}^{m}\right|-1},$$ 
where $S_{n}^{m}$ is the $n$-th data sample in the VDT $m$. $D_{DT}^{m}$ represents the dataset of VDT m .These total number of VDT $|\mathcal{M}|$ vehicle’s digital twins are distributed among the cloud, i.e.,
\begin{equation*} 
  D_{DT}^{m_1} \cap D_{DT}^{m_2} = \emptyset, \forall m_1,m_2\in \mathcal{M}, m_1 \ne m_2.  
\end{equation*}
The above equation explains that the dataset intersection of VDT $m_1$ and $m_2$ is an empty set where $m_1$ and $m_2$ VDT are part of VDT set M and $m_1$ is not equal to $m_2$. In simpler terms, it means VDT data elements from set M do not share common data, and set M has a unique set of elements. 
The priority of a VDT  is calculated as:
$$P_{DT} = \alpha_{cat} * DT_{cat},$$
where $ DT_{cat}$ is the category of a VDT, and $\alpha_{cat}$ is the weight associated with this category(higher for more priority). If there is more than one VDT with the same priority, then we give preference according to the category and state of their applications, for instance, active road and safety applications as shown in \textbf{Table \ref{tab:table3}}.
The priority of the application can be calculated as:
\begin{equation*} 
  P_{App} = \beta_{cat} * App_{cat}
\end{equation*}
where $ App_{cat}$ is the category of an application, and $\beta_{cat}$ is the weight associated with this application (higher for more critical applications) according to its category.

So, the total priority for resource acquisition is 
$$resource\_ acquision\_ priority(P) = P_{DT} + P_{App}$$
which explains that while assigning resources if the priority because of vehicle category is the same then the priority according to application category will be considered.  
In problem formulation, we introduce some specific requirements that apply to VDTN and define the VDT model with the following requirements satisfied:
\begin{enumerate}
\item A base station can concurrently connect to multiple vehicles, but a vehicle only connects to a particular base station;
\item Each vehicle has only one unique digital twin representation in the cloud;
\item Each digital twin connects with at least one other digital twin in the cloud to share or process data;
\item The communication links are \textit{intra-twin} and \textit{inter-twin} communication to communicate between vehicles, VDTs, and vehicles-VDTs as well;
\item Limited storage and computing capability should be available in each vehicle for temporary storing and computing limited independent and identically distributed data;
\item  For extensive computation or sharing with massive communication overhead, we use priorities for VDTs on the cloud. 
\end{enumerate}
\section{Priority-based  inter-twin communication in VDTNs} \label{sec:priority_based_comm}
In this section, we describe the proposed Priority-based inter-twin communication in vehicular digital twin architecture including inter-twin communication and intra-twin communication. The use of priority for managing communication overhead for emergency-providing vehicles’ digital twins is valuable in reducing latency and delays that occur in inter-twin communication of digital twins. 
\subsection{Intra-twin communication}
Figure 1 shows the architecture of the proposed Priority inter-twin communication in VDT architecture, which consists of a physical and digital twin Layer, base stations, and a set of vehicles digital twins $|\mathcal{M}|$. In intra-twin communication, communication takes place between the vehicles and their equivalent VDTs, which reside in the cloud. The coordination mechanism exploits intra-twin communication as specified in the algorithm, which generally consists of the following phases: 
\begin{enumerate}
\item{\textbf{Digital Twins Data Upload}}: The vehicle uploads the updated and real-time sensing data through intra-twin communication to its digital twin which is on the cloud side. They upload the data using wireless communication, for instance, 5G cellular network, WiFi, etc.;
\item{\textbf{Updated Information Transmission}} 
When private data is uploaded, it is processed at the digital twin layer. Suppose there is a total of k data updates using the Inter-Twin Algorithm.
\item{\textbf{Final Information Update}}
It can be known from the Intra-Twin Algorithm that after k times of Inter-Twin update, the finalized updated private data is obtained and transmitted back to the corresponding vehicle to proceed.
\end{enumerate}
Some characteristics are associated with Intra-twin communication which are discussed below:
\begin{itemize}
\item{\textbf{Real-time Simultaneity}}: Real-time simultaneity is required between autonomous vehicles and digital twins so that secure, accurate,  data is maintained, which meets the requirement. It also ensures a good user experience.
\begin{algorithm}[H]
\caption{Intra-Twin Communication Update}\label{alg:alg1}
\begin{algorithmic}
\STATE 1. Initialize Data Upload:
\STATE \hspace{\algorithmicindent} a) Each vehicle $m$ carries certain independent private data, $PD$, related to the vehicle;
\STATE \hspace{\algorithmicindent} b) It uses existing networks, base stations, and wireless communications (5G, WiFi), in this regard, which resides at the physical layer.

\STATE 2. Data Processing at Digital Twin Layer\\
\textbf{for} each update $k = 0,1,2, \cdots ,k – 1$ \textbf{do}\\
\quad\quad \textbf{\textsc{INTER-TWIN}}(${PD}$);
\STATE 3. Obtain the final updated private information, $PI$, from the digital twin layer to the base station, and transmit it back to the corresponding vehicle.

\end{algorithmic}
\label{alg1}
\end{algorithm}

\item{\textbf{Reliable Communication}} To maintain information reliability autonomous vehicles will have to maintain uninterrupted communication with the digital twins. In this way, updated data about the location and state of the vehicle will be concurrently uploaded to its corresponding digital twin, which might change as the vehicle moves on.
\item{\textbf{Connected and Confidential Connection}}: As the digital representation of the autonomous vehicle, the digital twin of the autonomous vehicle's objective is to use the cloud virtual resources. So, each digital twin is confidential to its corresponding vehicle and cannot be used by other vehicles. So, we shall have to maintain the security and privacy of the communication connection between them.   
\end{itemize}
\subsection{Inter-twin priority communication}
\textbf{Figure 1} shows the workflow of Inter-twin communication with Intra-twin communication which enables cooperation between the physical layer and digital twin layer. VDTN architecture that represents each vehicle through its corresponding digital twin in the cloud. The vehicle's digital twins can be divided into three categories according to the priority of the vehicle: High Priority VDT (HPVDT), Medium Priority VDT (MPVDT), and Normal Priority VDT (NPVDT). The vehicle with emergency services is referred to as a High Priority Vehicle, the vehicle with an easy passing route without legal requirements is referred to as a Medium Priority Vehicle, and the vehicle with no urgency features is referred to as a Normal Priority Vehicle. In the same way, if the same category VDTs are trying to access limited resources then priority would be given according to the application category. The detailed definitions of each of these categories are explained and shown in \textbf{Table \ref{tab:table2}} and \textbf{Table \ref{tab:table3}}.
\begin{table}[!t]
\caption{Brief Description of Categories Related to Vehicles \label{tab:table2}}
\centering
\begin{tabularx}{\textwidth}{|c||X|}
\hline
\textbf{Vehicle Priority} & \textbf{ Categories of Vehicles}\\
\hline
High Priority Vehicles & Police, Army, Customs, Ambulances, Law Enforcement Agencies, Fire Departments, Mobile Medical Units, and Detainees Vehicles.\\
\hline
Medium Priority Vehicles & Medical Association, Cash transportation, Blood Products contain vehicles, Human organs vehicles, and sanitary transport vehicles. \\
\hline
Normal Priority Vehicles & Private Vehicles, Vans, Goods Transport, Passenger Carrying vehicle.\\
\hline
\end{tabularx}
\end{table}

\begin{table}[!t]
\caption{Brief Description of Categories Related to Vehicles Applications\label{tab:table3}}
\label{tab:table1}
\centering
\begin{tabular}{|c||c|}
\hline
\textbf{Application Priority} & \textbf{ Categories of Applications}\\
\hline
First Priority Applications & Safety Applications\\
\hline
Second Priority Applications & Traffic Management Applications\\
\hline
Third Priority Applications & Efficiency Applications\\
\hline
Fourth Priority Applications & Information and Entertainment Applications\\
\hline
Fifth Priority Applications & Social Applications\\
\hline
\end{tabular}
\end{table}

The complete workflow and detailed definitions of these three types of vehicle digital twins and their communication with the physical layer through the base station are described as follows (for the workflow also see Algorithm 2):

Assumption 1: The inter-twin communication takes place in the cloud, which means that AI learning and data sharing, are among digital twins in the cloud.
\begin{enumerate}
\item{\textbf{Base Station: }}
Base station B first forwards the data to the digital twin of the corresponding vehicle on the digital twin layer. Whenever the digital twins complete their processing, they send the result to the base station, and the base station then forwards the result to the vehicle. 
\item{\textbf{High Priority Vehicle:}}
High Priority Vehicle $V_{high}$ consists of emergency services providing vehicles which include fire departments, customs, police, ambulances, army, mobile medical units, and detainee vehicles. The majority of these vehicles use sound to intimate their emergency services. These vehicles require a response from the network as early as possible which might cause a delay in the other vehicle's response accessing the same network and data resources. This is very much acceptable due to the emergency objective they are achieving. After the processing is done the other vehicles may continue with the same speed. High Priority Vehicle private data will be handled by its respective high-priority DT.
\item{\textbf{Medium Priority Vehicle:}}
Medium Priority Vehicle $V_{medium}$ symbolizes the vehicles that can have limited response delay privileges. So, it is not necessary to get an immediate response straight away and start processing their data but it is recommended to keep the delay limited. The vehicles that are part of this category include Blood products, cash transportation, human organs, and sanitary, and medical association vehicles. Medium-Priority Vehicle private data will be handled by its respective medium-priority DT.
\item{\textbf{Normal Priority Vehicle:}}
Normal Priority Vehicle $V_{normal}$ consists of vehicles with no emergency and unique needs. It includes vehicles that are private, personal, Passenger, and Goods-carrying vehicles. Normal-Priority Vehicle private data will be handled by its respective normal-priority DT. So, in case of congestion, sharing, or processing of data, $V_{high}$ vehicles will be treated first, $V_{medium}$ secondly, and $V_{normal}$ in the end.
\item{\textbf{Application Priority:}}
In the case of AppPriority() if two vehicles have the same priority and they are accessing the same resources in case of congestion, then the vehicles that are running safety applications will be given resources first, traffic management will be given second, efficiency apps will be given third, the infotainment application will be considered forth and the Social Applications application will be considered last in the end as shown in \textbf{Table \ref{tab:table3}}. If the application priority is also the same then the \textit{First Come First Serve} (FCFS) algorithm will be implemented according to the priority of the vehicle.
\item{\textbf{Aging:}}
We would also use the concept of Aging to gradually increase the priority of a waiting low-priority process of a vehicle to eliminate starvation. 
\item{\textbf{Indirect Result Update:}}
The autonomous vehicle in this way can indirectly communicate with other vehicles. The benefit of inter-twin communication is now vehicles can now communicate outside of their communication range with other vehicles. Another benefit is that now we can get global information that is not bound to a specific area or region. 
\end{enumerate}

\begin{algorithm}[H]
{\small 
\caption{Inter-Twin Communication Priority Update}\label{alg:alg1}
 \textbf{Input:} The data that needs to be processed from the base station \textbf{B} to the respective digital twin (\textbf{DT});\\
 \textbf{Output:} The processed information from the respective digital twin (\textbf{DT}) to the base station \textbf{B};
\begin{algorithmic}
\STATE\textbf{\textsc{INTER-TWIN}}$(PD):$
\STATE \textbf{(I). For high-priority vehicle DT:}
\STATE \textbf{Receive \textit{PD} from the base station;}
\STATE \textbf{Record the resources it requires into collection $Resource_{high}$};
\STATE \textbf{for} \textbf{all resources} $r  \in$ $Resource_{high}$ \textbf{do}
\STATE\hspace{0.5cm} if $r \notin$ $DT_{high}$ then
\STATE\hspace{0.8cm} Send $PD$ to $r$;
\STATE\hspace{0.8cm} Receive $PD$ from $r$;
\STATE\hspace{0.8cm} Update $PD$; 
\STATE\hspace{0.5cm} else if
\STATE\hspace{0.8cm} AppPriority();
\STATE\hspace{0.5cm} else
\STATE\hspace{0.8cm} FCFS();
\STATE\hspace{0.5cm} end if
\STATE end for
\STATE Send PD to the base station;
\STATE
\STATE(II).\textbf{ For medium-priority vehicle DT:}
\STATE \textbf{Receive \textit{PD} from the base station;}
\STATE \textbf{Record the resources it requires into collection} $Resource_{medium}$;
\STATE \textbf{for} all resources $r  \in$ $Resource_{medium}$ \textbf{do}
\STATE\hspace{0.5cm} if $r \notin$ $DT_{medium}$ and $r \notin$ $DT_{high}$ then
\STATE\hspace{0.8cm} Send $PD$ to $r$;
\STATE\hspace{0.8cm} Receive $PD$ from $r$;
\STATE\hspace{0.8cm} Update $PD$; 
\STATE\hspace{0.5cm} else if $r \in$ $DT_{medium}$
\STATE\hspace{0.8cm} AppPriority();
\STATE\hspace{0.5cm} else
\STATE\hspace{0.8cm} FCFS();
\STATE\hspace{0.5cm} end if
\STATE end for
\STATE Send $PD$ to the base station;
\STATE
\STATE(III).\textbf{ For normal-priority vehicle DT:}
\STATE \textbf{Receive \textit{PD} from the base station;}
\STATE \textbf{Record the resources it requires into collection} $Resource_{normal}$;
\STATE \textbf{for} all resources  $r \in 
 Resource_{normal}$ \textbf{do}
\STATE\hspace{0.5cm} if $r \notin$ $DT_{normal}$, $r \notin$ $DT_{medium}$ and $r \notin$ $DT_{high}$ then
\STATE\hspace{0.8cm} Send $PD$ to $r$;
\STATE\hspace{0.8cm} Receive $PD$ from $r$;
\STATE\hspace{0.8cm} Update $PD$; 
\STATE\hspace{0.5cm} else if  $r \in$ $DT_{normal}$
\STATE\hspace{0.8cm} AppPriority();
\STATE\hspace{0.5cm} else
\STATE\hspace{0.8cm} FCFS();
\STATE\hspace{0.5cm} end if
\STATE end for
\STATE Send $PD$ to the base station;
\end{algorithmic}
\label{alg1}
}
\end{algorithm}

The inter-twin communication can be characterized by the following aspects: 
\begin{itemize}
\item{\textbf{Peer-to-peer distributed connection}}

Each digital twin functions as an independent entity and multiple digital twins form a distributed network in inter-twin priority communication. Each digital twin entity in this network shares resources such as content or services following the priority of the digital twin. Consequently, every digital twin acts as both a resource acquirer and a resource provider.
\item{\textbf{Confidential data assets}}

The digital twins are isolated to their autonomous vehicles. They also store massive amounts of data that are derived from different operations of the vehicles. These data possess personal information of the vehicle and users and have a particular value. So, they are confidential data assists of the digital twins. 
\item{\textbf{Communication among multiple agents}}

Digital twins can be considered, as agents because of their autonomous virtual entity and intelligence. Consequently, the communication between the digital twins is considered, as a form of multi-agent communication.
\end{itemize}

\section{Result and Evaluation} \label{sec:eval}
This section includes a thorough evaluation of our suggested priority-based model's high usefulness and efficiency in performance metrics tasks under a range of dynamic topologies and data distributions.  

\subsection{Experimental Setup}
JGraphT\footnote{A network structure package for Java (https://jgrapht.org/)} creates the random cluster topologies and PriorityQueue class from the java.util package \footnote{The collections framework package for Java (https:// docs.oracle.com/javase/8/docs/api/java/util/package-summary.html)}implements the algorithms framework. 
This experimental setup focuses on scheduling based on the priority of the incoming task. To analyze the priority, round-robin\cite{abraham2006os}, and throttled load balancing algorithm\cite{hussein2015throttled} as explained in Baseline Studies, we have CloudSim 6.0 and a tool based on CloudSim \cite{priority}. CloudSim serves as an autonomous platform designed for modeling data centers, service intermediaries, and the scheduling and allocation policies of expansive cloud platforms. We use it to implement our inter-communication algorithm and load balancing on the cloud side of Digital Twins. The VM load balancing algorithms and our priority algorithm are implemented using the Java programming language.
To assess different load balancing policies, adjustments must be made to the settings of various components. We have configured parameters for application deployment, data center setup, and user base configuration as part of this implementation.

We assume that the application has been deployed within two data centers with ten virtual machines (VMs), each equipped with 1 GB of memory, running on a physical host with a capacity of 1000 MIPS. The specifics of the data center configuration, the configurations of the simulator, and the simulator setup values are detailed in TABLE \ref{tab:table4}. 

\begin{table}[!t]
\caption{\textbf{Data Center Setting}\label{tab:table4}}
\centering
\begin{tabular}{|c||c|}
\hline
\textbf{Parameter} & \textbf{Value}\\
\hline
Data Center OS & Linux\\
\hline
VM Memory & 1 GB \\
\hline
Data Centre Architecture & XEN\\
\hline
VM Bandwidth & 1000 \\
\hline
\end{tabular}
\end{table}

\subsection{Setting for Cluster}
Ten clusters are formed each containing ten VDT. Each cluster may consist of high-priority VDT, medium-priority VDT, or normal-priority VDT as represented by vehicles. Ten network topologies mentioned above each containing ten VDT, were randomly generated to simulate the dynamic network topology of multi-hop cluster VDT. 
\subsection{Datasets}  
  We conducted experiments using an actual driving dataset by vehicular mobility trace \footnote{A in CSV format, zipped (https://vehicular-mobility-trace.github.io/index.html\#dataset)}. The dataset contains real-time data about vehicles like the vehicle angle, vehicle position, vehicle type, two-dimensional vehicle coordinates, Vehicle ID, Speed, etc. in CSV format. As part of step Preprossing, we did data cleaning and normalized the data using StandardScalar.We randomly assigned high, medium, and normal priority to these vehicles by assigning weight to the vehicles as mentioned in the problem formulation. More weight is represented by less number for instance number one priority is more than number 2. Using this data, each node created in the cloud represents a VDT of its respective vehicle.
\subsection{Baseline Studies} 
We will use this data to do a comparison between the existing round-robin algorithm, and throttled load balancing algorithm with our priority algorithm. We have used the round-robin algorithm because it has no starvation issues, every job gets a fair allocation of CPU and in average response time, it has good performance. On the other hand, the throttled load balancing algorithm improves data center environment performance. It efficiently manages the loads of the virtual machines.
\subsection{Performance Metrics} 
We calculate message delivery rate, latency, Bandwidth Utilization, Throughput, Response Time, Resource Utilization, Fairness, Communication Overhead, Adaptability to Dynamic Changes, Fault tolerance, and Algorithm Complexity to do the comparison. Message delivery rate indicates the rate at which messages are delivered by the system, latency represents delay, Bandwidth Utilization represents the percentage of time the communication channel is utilized, throughput represents the rate of successful message delivery, response time indicates the average time taken to respond, resource utilization the percentage of time the vehicles are busy communicating, fairness means the degree to which the workload is divided equally among the servers, communication overhead means the amount of data exchanged between vehicles, Adaptability to Dynamic Changes means algorithm's ability to adjust to variations in the environment, Fault tolerance means how well algorithm handle VDT failures in the environment. Algorithm Complexity means the amount of computational overhead needed by the algorithm.

\begin{figure*}[!ht]
    \centering
    \includegraphics[width=5.6 in]{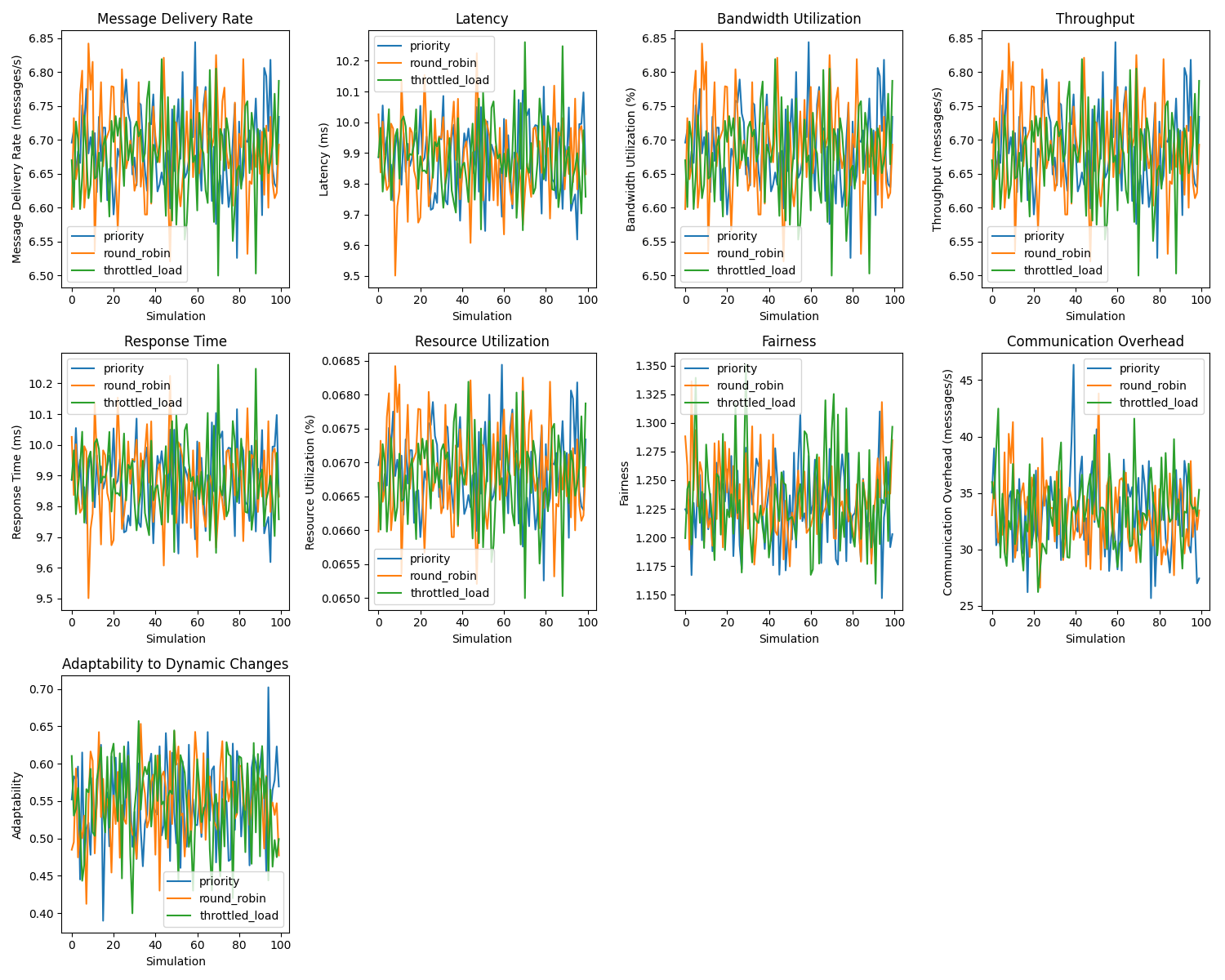}
    \caption{Performance metrics of our proposed priority algorithm with round robin and throttled load balancing algorithm}
    \label{fig_2}
\end{figure*}

\subsection{Experiment Results and Performance Analysis}
We have implemented our priority VDT algorithm and conducted an analysis using the CloudSim tool. This implementation involved extending the existing classes within the simulation tool. We have imported the real-time vehicular mobility trace CSV file dataset. The following analyses encompassed average performance metrics for 100 simulations of VDT. The comparative analysis of the average result of performance metrics for all the algorithms is presented in TABLE \ref{tab:table5}.
\begin{table*}[!t]
\captionsetup{justification=centering}
\caption{\textbf{Experimental results regarding the average performance metrics across a couple of algorithms with Priority Algorithm}\label{tab:table5}}
{\footnotesize 
\setlength{\tabcolsep}{4pt} 
\begin{tabularx}{1.15\textwidth}{|c|*{11}{X|}}
\hline
\textbf{Algorithm} & \textbf{MDR}& \textbf{LAT}& \textbf{BU}& \textbf{TP}& \textbf{RT}& \textbf{RU}& \textbf{FAIR}& \textbf{CO}& \textbf{ADC}& \textbf{ FT}& \textbf{AC}\\
\hline
Round Robin & 6.6881& 9.8850& 6.6882&6.6884 &9.8852 &0.0667 &1.227 & 33.2&0.546 & 0.00&14.0\\
\hline
Throttled Load balancing algorithm & 6.6770&9.8849&6.6771 &6.6773 &9.8851 &0.0666 &1.232 &33.3 & 0.547& 0.00&11.0\\
\hline
Priority Based VDTN algorithm & 6.6883&9.8830&6.6884& 6.6886& 9.8832&0.0668& 1.233& 33.0&0.549 &0.00 &8.0\\
\hline
\end{tabularx}
} 
\end{table*}
\footnote{MDR = Message Delivery Rate (messages), LAT= Latency(ms), BU = Bandwidth Utilization( \%),TP = Throughput(messages/s),RT = Response Time(ms),RU = Resource Utilization(\%),FAIR = Fairness,CO = Communication Overhead (messages/s),ADC = Adaptability to Dynamic Changes, FT = Fault Tolerance, AC = Algorithm Complexity}

Figure 2 illustrates the graph related to the performance metrics comparison of the round-robin, throttled algorithm, and priority-based VDTN.  It is worth noting from TABLE \ref{tab:table5} that our proposed algorithm produces better values for the performance metrics of Message Delivery Rate(MDR), Bandwidth Utilization(BU), Throughput(TP), Resource Utilization(RU), Fairness and Adaptability to Dynamic Changes(ADC) compared to the existing throttled load balancing and round-robin algorithm and our proposed algorithm produces low values for Latency, Response Time(RT), Communication Overhead(CO) and Algorithm Complexity(AC) which confirm our algorithm efficiency over them. However, the Fault Tolerance value remains zero for all three algorithms because once there is any fault all three algorithms behave in the same way. 

Overall, the suggested priority algorithm works better than previous algorithms with more efficiency as well as effectiveness and less delay and communication overhead, as shown by the numerical data in TABLE \ref{tab:table5}.

\begin{figure*}[!ht]
    \centering
    \includegraphics[width=5.6in]{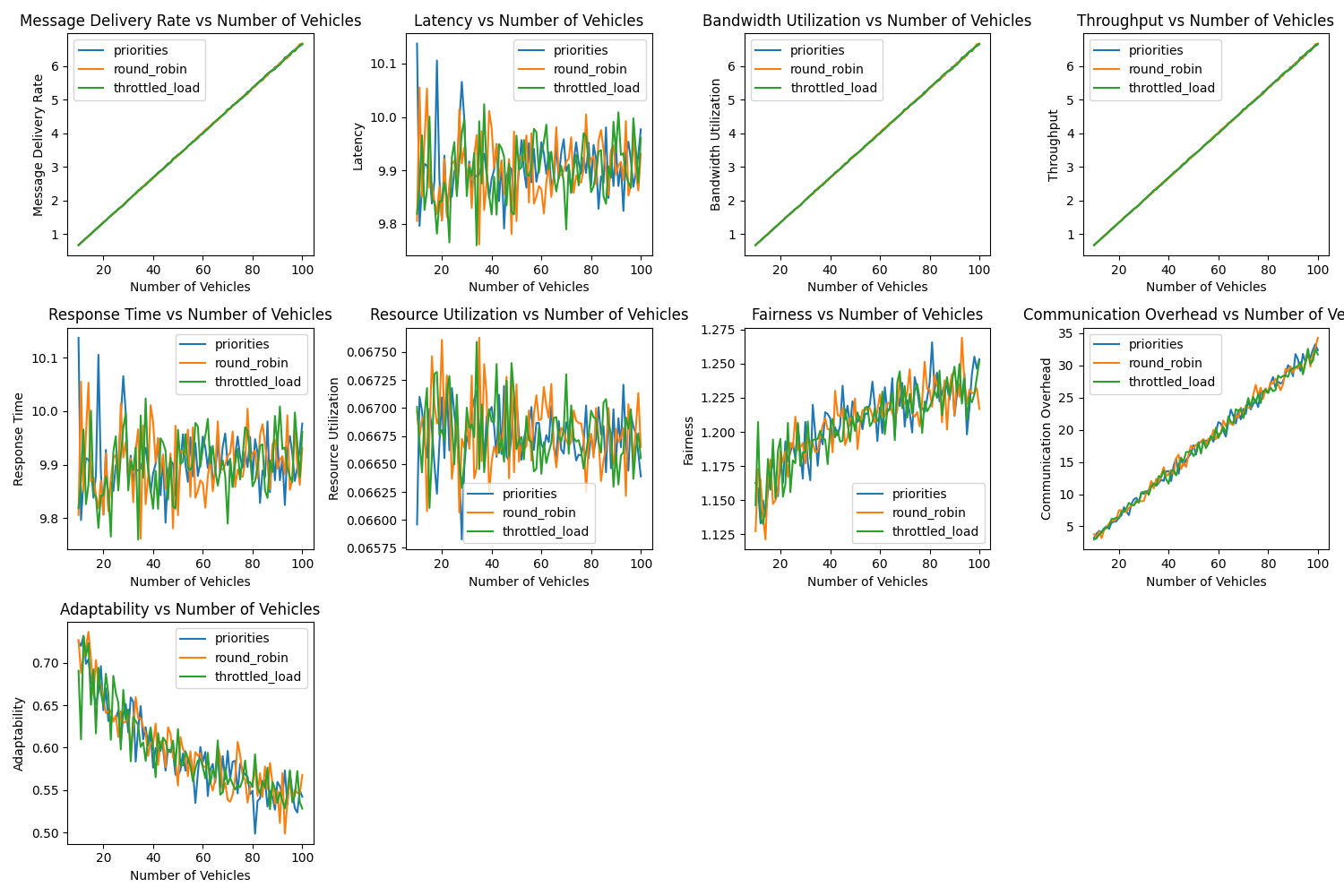}
    \caption{Scalability of our proposed priority algorithm with round robin and throttled load balancing algorithm}
    \label{fig_3}
\end{figure*}

Figure 3 shows the scalability performance of all the algorithms. The proposed priority algorithm performed better when compared to the round-robin and throttled load balancing algorithm as we scaled our network in the dynamic environment.  As we scale our network, the Fault Tolerance(FT) would still perform the same way for all three algorithms. However, the Algorithm Complexity(AC) of our algorithm would remain better as compared to the other two. 

The results based on the calculation of average performance metrics for different categories of algorithms are shown in TABLE \ref{tab:table6}. 
\begin{table*}[!t]
\captionsetup{justification=centering}
\caption{\textbf{Experimental results regarding the average scalability metrics across a couple of algorithms with Priority Algorithm}\label{tab:table6}}
{\footnotesize 
\setlength{\tabcolsep}{4pt} 
\begin{tabularx}{1.15\textwidth}{|c|c|*{11}{X|}}
\hline
\textbf{Algorithm} & \textbf{SV} & \textbf{MDR}& \textbf{LAT}& \textbf{BU}& \textbf{TP}& \textbf{RT}& \textbf{RU}& \textbf{FAIR}& \textbf{CO}& \textbf{ADC}& \textbf{FT}& \textbf{AC}\\
\hline
\multirow{3}{*}{Round Robin} &10 & 0.665& 9.9& 0.666&0.668 &9.7 &0.069 &1.130 & 3.4&0.72 & 0.00&11.0\\
\cline{2-13}
& 20& 1.338&9.85&1.339 &1.340 &9.88 &0.069 &1.16 &7.3 & 0.64& 0.00&11.0\\
\cline{2-13}
&30 & 1.998&9.90&1.94& 1.96& 9.87&0.066& 1.17& 10.2&0.651 &0.00 &11.0\\
\hline
\hline
\multirow{3}{*}{Throttled Load balancing algo} &10 & 0.651& 9.8& 0.670&0.672 &9.6 &0.073 &1.139 & 3.2&0.71 & 0.00&14.0\\
\cline{2-13}
& 20& 1.346&9.86&1.347 &1.350 &9.80 &0.073 &1.18 &7.0 & 0.65& 0.00&14.0\\
\cline{2-13}
&30 & 1.990&9.98&1.93& 1.94& 9.88&0.063& 1.18& 9.8&0.619 &0.00 &14.0\\
\hline\hline
\multirow{3}{*}{Priority Based VDTN algo} &10 & 0.672& 9.7& 0.673&0.675 &9.5 &0.076 &1.151 & 3.1&0.74 & 0.00&10.0\\
\cline{2-13}
& 20& 1.349&9.76&1.350 &1.352 &9.64 &0.074 &1.19 &6.6 & 0.66&0.00&10.0\\
\cline{2-13}
&30 & 2.005&9.86&2.00& 1.99& 9.84&0.068& 1.19& 9.4&0.656 &0.00 &10.0\\
\hline
\end{tabularx}
} 
\end{table*}
\footnote{SV = No of Vehicles Scale,MDR = Message Delivery Rate (messages), LAT= Latency(ms), BU = Bandwidth Utilization( \%),TP = Throughput(messages/s),RT = Response Time(ms),RU = Resource Utilization(\%), FAIR= Fairness,CO = Communication Overhead (messages/s),ADC = Adaptability to Dynamic Changes, FT = Fault Tolerance, AC = Algorithm Complexity}
\section{Conclusion} \label{sec:concl}
Priority-based Vehicular Digital Twin Network is an essential model for transportation in the future. In this article, we have proposed an efficient and effective priority-based model for vehicular digital twin networks to reduce response time and delay especially for emergency vehicles. We first introduced the priority-based vehicular digital twin architecture then we described the related work literature of vehicular digital twin ad hoc networks. We then discuss the problem formulation section. In addition, two algorithms for intra-twin communication and inter-twin priority communication were designed respectively which can decrease the communication overhead for the emergency vehicles, our algorithm ensures that emergency vehicles services are communicated swiftly, limiting latency and raising the network overall consistency . Performance metrics experiments on real datasets are conducted with two existing algorithms demonstrating our proposed algorithm's effectiveness and efficiency. Additionally, we also proved our algorithm's effectiveness through scalability in dynamic environments.

The findings demonstrate that the priority-based algorithm greatly improves VDT responsiveness in situations where emergency vehicles require immediate response and communication.Deploying priorities in Vehicular Digital twin through inter-twin communication can greatly help reduce accidents due to delays in response, oversee traffic, provide a solution for vehicles, and improve reliability. 

Our work contributes a novel approach to the priority management issue in VDT which adds to the growing domain of intelligent transportation systems. The suggested algorithm creates an opportunity to introduce priorities in the Digital Twin of VANET in addition to communication efficiency.  

We expect this work to be beneficial for implementing priorities in vehicular digital twin networks. In the future, we will make more research efforts into introducing more vehicular scenarios, communication protocols, machine learning algorithms, blockchain technology, security, scalability, and privacy in this emerging field. We shall also try to extend our proposed priority-based Vehicular Ad-hoc Network model and its impact on the performance of more cooperative and autonomous driving applications of the real world.

In summary priority-based VDT is a promising development in intelligent transportation which would make VDT more safer and efficient.

\end{document}